\documentclass[12pt,a4paper]{article}
\usepackage{a4wide}
\usepackage[centertags]{amsmath}
\usepackage{amssymb}
\usepackage{epsfig}
\usepackage{cite}

\newcommand{\id}{1\!\!1}
\renewcommand{\thefootnote}{\fnsymbol{footnote}}

\begin{document}
\title{
\begin{flushright}
\ \\*[-80pt]
\begin{minipage}{0.2\linewidth}
\normalsize
%hep-th/0612044 \\
IPPP/06/86 \\
DCPT/06/172 \\
KUNS-2055 \\*[50pt]
\end{minipage}
\end{flushright}
{\Large \bf
Non-Factorisable \boldmath{${\mathbb Z}_2 \times {\mathbb Z}_2$}
Heterotic Orbifold Models 
and Yukawa Couplings
\\*[20pt]}}

\author{
Stefan~F\"orste$^{1,}$\footnote{E-mail
  address: stefan.forste@durham.ac.uk}, \ 
Tatsuo~Kobayashi$^{2,}\setcounter{footnote}{3}$\footnote{
E-mail address: kobayash@gauge.scphys.kyoto-u.ac.jp}, \
Hiroshi~Ohki$^{3,}$\footnote{
E-mail address: ohki@scphys.kyoto-u.ac.jp} \\
\ and \
Kei-jiro Takahashi$^{2,}$\footnote{
E-mail address: keijiro@gauge.scphys.kyoto-u.ac.jp}\\*[20pt]
$^1${\it \normalsize
Institute for Particle Physics Phenomenology, University of Durham,
South Road, }\\ {\it \normalsize Durham DH1 3LE, UK} \\
$^2${\it \normalsize
Department of Physics, Kyoto University,
Kyoto 606-8502, Japan} \\
$^3${\it \normalsize
Department of Physics, Kyoto University,
Kyoto 606-8501, Japan} \\*[50pt]}

\date{
\centerline{\small \bf Abstract}
\begin{minipage}{0.9\linewidth}
\medskip
\medskip
\small
We classify compactification lattices for supersymmetric ${\mathbb
  Z}_2 \times {\mathbb Z}_2$ orbifolds. These lattices include
factorisable as well as non-factorisable six-tori. Different models
lead to different numbers of fixed points/tori. A lower bound on the
number of fixed tori per twisted sector is given by four, whereas an
upper bound consists of 16 fixed tori per twisted sector. 
Thus, these models have a variety of generation numbers.
For example, in the standard embedding, the smallest number of 
net generations among these classes of models is equal to six, 
while the largest number is 48.
Conditions for allowed Wilson lines and Yukawa couplings are derived. 
\end{minipage}
}

\begin{titlepage}
\maketitle
\thispagestyle{empty}
\end{titlepage}

\renewcommand{\thefootnote}{\arabic{footnote}}
\setcounter{footnote}{0}

\section{Introduction}

The Theory of Superstrings successfully unifies the concepts of
Quantum Field Theory and General Relativity. In order to claim that
it incorporates a unification of all forces observed in nature, one
has to prove the existence of string models reproducing standard
particle physics. The best way to prove that existence consists in the
construction of explicit models since that allows also to investigate
phenomenological implications of string theory. The, perhaps most
traditional, attempt of identifying realistic string models is given by
heterotic orbifold constructions  \cite{Dixon,IMNQ}.
In more recent years, this line of research was boosted by the
observation that phenomenological properties can be connected to
geometrical properties of the orbifold
\cite{Kobayashi:2004ud,Forste:2004ie,Buchmuller:2004hv}. 
Examples for quantities which
are directly tied to geometry are the K\"ahler potential for twisted
sector states as well as Yukawa couplings \cite{Hamidi:1986vh,Dixon:1986qv,
Burwick:1990tu,Kobayashi:2003vi,Kobayashi:2003gf}. (For interesting
applications see e.g.\ Ref.\ \cite{Ko:2004ic}.)

In a parallel development, semi realistic models have been obtained in
the free fermionic formulation of heterotic strings
\cite{Antoniadis:1989zy,Faraggi:1991jr}. Although there are some
indications \cite{Faraggi:1993pr} that these models are related to
${\mathbb Z}_2 \times {\mathbb Z}_2$ orbifolds a precise connection
has not been worked out in general. Hence a geometric picture is
missing for many free fermionic models. 

Most of the earlier explicit ${\mathbb Z}_2 \times {\mathbb Z}_2$
heterotic constructions  are based on factorisable compactification
lattices \cite{Font:1988mk,Forste:2004ie} (for different
compactifications for this and other orbifolds see, however,
\cite{Dixon, Dixon:1986yc, Bailin:1994ma, Donagi:2004ht}). In the
current paper, we are going to further explore the approach taken in
\cite{Faraggi:2006bs} where the explicit construction on ${\mathbb Z}_2
\times {\mathbb Z}_2$ orbifolds of non factorisable six-tori with and
without Wilson lines is presented. These types of orbifolds are among
the simplest phenomenologically interesting constructions of string
theory models. Since it is believed, that string theory allows for a
large number of realistic models, simplicity provides a good
principle to impose. Moreover, the phenomenologically attractive
features of the previously mentioned free fermionic constructions and
their conjectured relation to ${\mathbb Z}_2 \times {\mathbb Z}_2$
orbifolds suggest a potential relevance of ${\mathbb Z}_2 \times
{\mathbb Z}_2$ for real particle physics. 

Replacing factorisable compactification lattices by non factorisable
ones affects the fixed point structure of the internal space. In
${\mathbb Z}_2 \times {\mathbb Z}_2$ chiral matter originates from
twisted sector states and, hence, depends on the fixed point
structure. But also other relevant aspects like the selection rules
\cite{Hamidi:1986vh,Font:1988mm} for Yukawa couplings are affected.
On the factorisable ${\mathbb Z}_2 \times {\mathbb  Z}_2$ orbifold,
only `diagonal' Yukawa couplings are allowed.
Here, the phrase ``diagonal Yukawa couplings'' means that
when we choose two fields, the other field to be allowed to couple
is uniquely fixed. 
Furthermore, allowed Yukawa couplings are of $O(1)$.
The situation will change on non-factorisable ${\mathbb Z}_2 \times
{\mathbb Z}_2$ orbifold models.

In the present paper, we will construct eight classes of ${\mathbb
  Z}_2 \times {\mathbb Z}_2$ orbifolds. All of these classes lead to
  $N=1$ supersymmetry in four dimensions. Topological properties of
  the orbifold differ from class to class but do not change for models 
within the same class. 

The paper is organised as follows. In the next section we discuss some
$T^3 / {\mathbb Z}_2 \times {\mathbb Z}_2$ orbifolds, and specify
under which circumstances two orbifolds can be considered as
equivalent. In section 3 we identify building blocks for the
classification of $T^6 / {\mathbb Z}_2 \times {\mathbb Z}_2$
orbifolds. Section 4 describes how to compute the number of families
and anti-families from modular invariance. In section 5 we discuss
consistency conditions on discrete Wilson lines and selection rules
for Yukawa couplings. We summarise and discuss our results in section
6. Appendix A provides the details for eight classes of $T^6/ {\mathbb
  Z}_2 \times {\mathbb Z}_2$ orbifolds.

\section{\boldmath{${\mathbb Z}_2 \times {\mathbb Z}_2$} Orbifolds of
  \boldmath{$T^3$}} 

As a warm up, we are going to discuss orbifolds of $T^3$. These will
serve as easy examples to identify equivalent orbifold
compactifications and also appear as substructures of the relevant
orbifolds of $T^6$ to be studied afterwards.  We will call orbifolds
equivalent if they are connected by a continuous deformation of
geometric moduli. From a phenomenological perspective, we would call
two orbifolds equivalent if the massless spectrum and the selection
rules for allowed couplings are the same. The latter understanding of
equivalence follows from our definition.  
In the following, we present two pairs of equivalent orbifolds of
$T^3$, and address both of the above aspects of equivalence.  

\subsection{ SU(4)/\boldmath{${\mathbb Z}_2 \times {\mathbb Z}_2$}
  versus SU(3)\boldmath{$\times$}SU(2)/\boldmath{${\mathbb Z}_2 \times
  {\mathbb Z}_2$} }

We specify the orbifold action on a Cartesian coordinate system of a
three dimensional Euclidean space by associating the generators of
${\mathbb Z}_2 \times {\mathbb Z}_2$, $\theta_1$ and $\theta_2$,  with 
three by three transformation matrices. In the present example, we choose
\begin{equation}
\theta_1 = diag\left( -1,1,1\right) \,\,\, ,\,\,\, \theta_2 = -\id_3 .
\end{equation}
The third non trivial element is given by the product $\theta_3 =
\theta_1\theta_2$. 
The three dimensional space is compactified by identifying points
differing by vectors of a three dimensional lattice. We take the SU(4)
root lattice  $\Lambda_{SU(4)}$ with the simple roots
\begin{equation} \label{eq:su4}
\alpha_1 = \left( \sqrt{2}, 0,0\right) \,\,\, , \,\,\, \alpha_2 =
\left( -\frac{\sqrt{2}}{2}, \sqrt{\frac{3}{2}} , 0\right) \,\,\, ,
\,\,\,
\alpha_3 = \left( 0, -\sqrt{ \frac{2}{3}}, \frac{2}{\sqrt{3}}\right) .
\end{equation}
On the lattice vectors, $\theta_1$ acts as
\begin{equation}
\theta_1 \alpha_1 = - \alpha_1, \qquad
\theta_1 \alpha_2 = \alpha_1 + \alpha_2, \qquad
\theta_1 \alpha_3 = \alpha_3 ,
\end{equation}
which is the Weyl reflection at $\alpha_1$. 

Next, we count the number of points/tori fixed under the action of the
non trivial orbifold elements $\theta_1$, $\theta_2$ and
$\theta_3$. We will employ a Lefshets fixed point theorem in order to
compute them. The element $\theta_1$ acts trivially on two directions
and hence leaves tori fixed. For the computation two lattices are
relevant.

The sublattice $(1-\theta_1)\Lambda_{SU(4)}$ is spanned by
$\alpha_1$.
In the non-factorisable orbifold, the lattice, which is normal to
the lattice invariant under the twist $\theta$, is non-trivial and
important.
We denote it by $N_\theta$.
The normal lattice $N_{\theta_1}$ is spanned by $\alpha_1$.
The number of independent fixed points/tori is obtained as
\cite{Narain:1986qm} (see also  \cite{Erler:1992ki,Faraggi:2006bs}) 
\begin{equation}
\frac{Vol\left((1-\theta_1)\Lambda_{SU(4)}\right)}
{Vol\left(N_{\theta_1}\right)},
\end{equation}
that is, the number of lattice sites on $N_{\theta_1}$,
which are not identified up to $(1-\theta_1)\Lambda_{SU(4)}$.
The $T_1$ sector\footnote{The $T_i$ sector corresponds to twisted
  strings closing only after points are identified by a $\theta_i$
  action.} has a single fixed torus corresponding 
to\footnote{Instead of giving the coordinates of the fixed torus we
  associate it to the space group element leaving it fixed also on the
  non compact three space. On the orbifold, this identifies the fixed
  torus uniquely.} $(\theta_1,0)$.
For the $T_2$ sector, the sublattice $(1-\theta_2)\Lambda_{SU(4)}$ is
spanned by $2\alpha_i$ $(i=1,2,3)$, and there are
eight fixed points corresponding to
$(\theta_2,\sum m^{(2)}_i\alpha_i)$ with $m^{(2)}_i=0,1$ for
$i=1,2,3$.
For the $T_3$ sector, the sublattice $(1-\theta_3)\Lambda_{SU(4)}$
is spanned by $(\alpha_1+2\alpha_2)$ and $2\alpha_3$, and
the normal lattice $N_{\theta_3}$ is spanned by
$(\alpha_1+2\alpha_2)$ and $\alpha_3$.
Thus, the $T_3$ sector has two independent fixed tori,
$(\theta_3,m^{(3)}_3\alpha_3)$ with $m^{(3)}_3=0,1$.

Now, let us present the orbifold of another $T^3$ which has the same
topology as the one discussed, so far. We choose the same
representation of the orbifold group on Cartesian coordinates but
replace the compactification lattice by the root lattice of
SU(3)$\times$SU(2), i.e.\ the lattice is generated by
\begin{equation}
\alpha_1 = \left( \sqrt{2}, 0,0\right) \,\,\, , \,\,\, \alpha_2 =
\left( -\frac{\sqrt{2}}{2}, \sqrt{\frac{3}{2}} , 0\right) \,\,\, ,
\,\,\,
e_3 = \left( 0, 0 , \sqrt{2}\right) .
\end{equation}
The action of $\theta_1$ on the basis of the lattice is  
\begin{equation}\label{eq:su2su3}
\theta_1 \alpha_1 = - \alpha_1, \qquad
\theta_1 \alpha_2 = \alpha_1 + \alpha_2, \qquad
\theta_1 e_3 = e_3 ,
\end{equation}
which is again a Weyl reflection at $\alpha_1$.

The sublattice $(1-\theta_1)\Lambda_{SU(3)\times SU(2)}$ is spanned by
$\alpha_1$, and the normal lattice $N_{\theta_1}$ is spanned by
$\alpha_1$.
Thus, the $T_1$ sector has a single independent fixed torus
$(\theta_1,0)$.
The sublattice $(1-\theta_2)\Lambda_{SU(3)\times SU(2)}$ is spanned by
$2\alpha_i$ $(i=1,2)$ and $2e_3$,
and the $T_2$ sector has eight fixed points,
$(\theta_2,\sum_i m^{(2)}_i\alpha_i + m^{(2)}_3 e_3)$.
The sublattice $(1 -\theta_3) \Lambda_{SU(3)\times SU(2)}$ is
spanned by $(\alpha_1+2\alpha_2)$ and $2e_3$,
and the normal lattice $N_{\theta_3}$ is spanned by
$(\alpha_1+2\alpha_2)$ and $e_3$.
There are two independent fixed tori $(\theta_3,m^{(3)}_3e_3)$
with $m^{(3)}_3=0,1$.

Now, let us compare the above two types of orbifolds.
Obviously, they have the same number of fixed points.
Moreover, the structure of $(1-\theta) \Lambda$ and $N_\theta$
is the same. That leads to the same selection rule of allowed couplings.
For the calculation, it is simpler to use a direct product of smaller
mutually orthogonal lattices. 

The above consideration indicates that, topologically, the two orbifolds
might be equivalent. In the following, we will see that they can
indeed be connected by continuous deformations of geometric moduli. In
Cartesian coordinates the geometric moduli of our orbifolds are the
following metric components
\begin{equation}
G_{11},\,\,\, G_{22},\,\,\,  G_{23}, \,\,\, G_{33} .
\end{equation}
Turning on these moduli changes the scalar products of lattice
vectors. For the system (\ref{eq:su4}) one obtains
\begin{equation}
\alpha_i \cdot \alpha_j = \left( \begin{array}{ c c c}
2 G_{11} & - G_{11} & 0 \\
- G_{11} & \frac{1}{2} G_{11} + \frac{3}{2} G_{22} & G_{23} - G_{22}
\\
0 & G_{23} - G_{22} & \frac{2}{3} G_{22} + \frac{4}{3} G_{33} -
\frac{4\sqrt{2} }{3} G_{23} 
\end{array} \right) .
\end{equation}
For the Euclidean metric one obtains the Cartan matrix of SU(4),
whereas one can find also a metric yielding the Cartan matrix of
SU(3)$\times$SU(2) 
\begin{equation}\label{eq:contdef}
\alpha_i \cdot \alpha_j = \left\{ \begin{array}{l c l}
\left( \begin{array}{ccc} 2 & -1 & 0 \\
                          -1&  2 & -1 \\
                           0 & -1 & 2 
\end{array} \right) & \mbox{for} & G_{ij} = \delta_{ij} \\
\\
\left(\begin{array}
{ccc} 2 & -1 & 0 \\
                          -1&  2 & 0 \\
                           0 & 0 & 2 
\end{array} \right) & \mbox{for} & \begin{array}{l}
G_{11}=G_{22}=G_{23}=1, \\
G_{33} = 1 + \sqrt{2} . \end{array} \end{array} \right.  
\end{equation}
The second metric in (\ref{eq:contdef}) can be transformed into a
Euclidean metric by coordinate changes. Since these leave the scalar
products invariant, one obtains an SU(3)$\times$SU(2) root lattice
embedded in Euclidean space. 

There is a, perhaps easier, way to see that the two orbifolds are
connected. Instead of varying the metric one could change the basis in
the 2-3 plane whilst keeping the metric fixed. Replacing $\alpha_3$ in
(\ref{eq:su4}) by $e_3$ in (\ref{eq:su2su3}) can be easily achieved by
such a change of basis.

\subsection{ SO(6)/\boldmath{${\mathbb Z}_2 \times {\mathbb Z}_2$}
  versus SO(4)\boldmath{$\times$}SU(2)/\boldmath{${\mathbb Z}_2 \times 
  {\mathbb Z}_2$} }

In the present example we represent the orbifold action on Cartesian
coordinates by
\begin{equation}\label{eq:orbidaccar}
\theta_1 = diag\left( 1,1,-1\right) \,\,\, ,\,\,\, \theta_2 = -\id_3 .
\end{equation}
As a compactification lattice we choose the SO(6) lattice with simple
roots
\begin{equation}\label{eq:so6}
\alpha_1 = \left( 0, 1, -1\right) \,\,\, , \,\,\, 
\alpha_2 = \left( 1, -1, 0\right) \,\,\, , \,\,\,
\alpha_3 = \left( 0, 1, 1\right) .
\end{equation}
With SU(4) being the universal covering of SO(6) this lattice is
equivalent to the previously considered lattice (\ref{eq:su4}). Our
terminology is inspired by the observation that the way roots have
been written fits into a general SU(N) respectively SO(2N) pattern. In
the present context, the real difference is that now $\theta_1$ acts
as an outer automorphism of the Dynkin diagram
\begin{equation}\label{eq:obact}
\theta_1 \alpha_1 = \alpha_3, \qquad \theta_1 \alpha_2 = \alpha_2,
\qquad \theta_1 \alpha_3 =\alpha_1.
\end{equation}
The sublattices $(1 - \theta_i)\Lambda_{SU(4)}$ and the
normal lattices $N_{\theta_i}$ are spanned by
\begin{equation}
\begin{array}{ll}
 (1 -\theta_1) \Lambda_{SU(4)}:\{ \alpha_1 - \alpha_3\}, &\qquad
N_{\theta_1}: \{\alpha_1 - \alpha_3\}, \\
 (1 -\theta_2) \Lambda_{SU(4)}:\{ 2\alpha_1,2\alpha_2,2\alpha_3 \},
& \qquad N_{\theta_2}: \{ \alpha_1,\alpha_2,\alpha_3 \},  \\
 (1-\theta_3)\Lambda_{SU(4)}:\{ (\alpha_1 + \alpha_3), 2\alpha_2 \},
& \qquad N_{\theta_3}: \{ (\alpha_1 + \alpha_3), \alpha_2 \}.
\end{array}
\end{equation}
The $T_1$, $T_2$ and $T_3$ sectors have one, eight and two fixed
points, respectively, which are denoted as
\begin{equation}
(\theta_1,0), \qquad (\theta_2,\sum_{i=1}^3 m^{(2)}_i \alpha_i),
\qquad (\theta_3,m^{(3)}_2 \alpha_2),
\end{equation}
where $m^{(2)}_i=0,1$ and $m^{(3)}_2 =0,1$.

Next, we are going to compare this to a compactification on an
SO(4)$\times$SU(2) lattice with the same action of the orbifold group
on Cartesian coordinates. The compactification lattice is (again the
terminology of writing SO(4) instead of SU(2)$^2$ is inspired by the
way roots $\alpha_1$ and $\alpha_3$ are written)
\begin{equation}\label{eq:so4tsu2}
\alpha_1 = \left( 0, 1, -1\right) \,\,\, , \,\,\, 
e_2 = \left( \sqrt{2}, 0, 0\right) \,\,\, , \,\,\,
\alpha_3 = \left( 0, 1, 1\right) .
\end{equation}
The action on these lattice vectors is the same as in (\ref{eq:obact})
with $\alpha_2$ replaced by $e_2$.

The sublattices $(1 - \theta_i)\Lambda_{SO(4)\times SU(2)}$ and the
normal lattices $N_{\theta_i}$ are spanned by
\begin{equation}
\begin{array}{ll}
 (1 -\theta_1) \Lambda_{SO(4)\times SU(2)}:\{ \alpha_1 - \alpha_3\}, &\qquad
N_{\theta_1}: \{\alpha_1 - \alpha_3\}, \\
 (1 -\theta_2) \Lambda_{SO(4)\times SU(2)}:\{ 2\alpha_1,2e_2,2\alpha_3 \},
& \qquad N_{\theta_2}: \{ \alpha_1,e_2,\alpha_3 \},  \\
 (1-\theta_3)\Lambda_{SO(4)\times SU(2)}:\{ (\alpha_1 + \alpha_3), 2e_2 \},
& \qquad N_{\theta_3}: \{ (\alpha_1 + \alpha_3), e_2 \}.
\end{array}
\end{equation}
The $T_1$, $T_2$ and $T_3$ sectors have one, eight and two fixed
points, respectively, which are denoted as
\begin{equation}
(\theta_1,0), \qquad (\theta_2,\sum_{i=1,3} m^{(2)}_i \alpha_i+m^{(2)}_2e_2 ),
\qquad (\theta_3,m^{(3)}_2 e_2),
\end{equation}
where $m^{(2)}_i=0,1$ and $m^{(3)}_2 =0,1$.

The last two orbifolds lead to the same number of fixed points.
Furthermore, the structure of $(1 - \theta ) \Lambda$
and $N_\theta$ is the same.
Thus, selection rules for allowed couplings impose the same conditions
for the two orbifolds.

Finally, we argue that the two orbifolds, SO(6)/${\mathbb Z}_2\times
{\mathbb Z}_2$ and SO(4)$\times$SU(2)/${\mathbb Z}_2\times
{\mathbb Z}_2$, can be continuously deformed
into each other by changing geometric moduli. 
The orbifold action (\ref{eq:orbidaccar}) leaves the metric components
\begin{equation}
G_{11},\,\,\, G_{12},\,\,\, G_{22},\,\,\, G_{33}
\end{equation}
invariant. In such general background the scalar product of the simple
SO(6) roots (\ref{eq:so6}) reeds
\begin{equation}
\alpha_i \cdot \alpha_j = \left( \begin{array}{c c c} 
G_{22} + G_{33} & G_{12} - G_{22} & G_{22} - G_{33}\\
G_{12} - G_{22} & G_{11} - 2 G_{12} + G_{22} & G_{12} - G_{22} \\
G_{22} - G_{33} & G_{12} - G_{22} & G_{22} + G_{33} \end{array}
\right) .
\end{equation}
There are special points in moduli space where the above becomes the
Cartan matrix of SO(6) or SO(4)$\times$SU(2),
\begin{equation}
\alpha_i \cdot \alpha_j = \left\{ \begin{array}{l c l}
\left( \begin{array}{ccc} 2 & -1 & 0 \\
                          -1&  2 & -1 \\
                           0 & -1 & 2 
\end{array} \right) & \mbox{for} & G_{ij} = \delta_{ij} \\
\\
\left(\begin{array}
{ccc} 2 & 0 & 0 \\
      0 &  2 & 0 \\
       0 & 0 & 2 
\end{array} \right) & \mbox{for} & \begin{array}{l}
G_{12}=G_{22}=G_{33}=1, \\
G_{11} = 3 . \\
\end{array} \end{array} \right.  
\end{equation}
Alternatively, one can deform geometric moduli by changing the basis
in the 1-2 plane while keeping the metric Euclidean. Such a basis change
can be employed to replace $\alpha_2$ in (\ref{eq:so6}) by $e_2$ in 
(\ref{eq:so4tsu2}).

Let us mention a subtlety which will be of some importance
later. Instead of (\ref{eq:so6}) we could have chosen the basis
\begin{equation}\label{eq:so6t}
\tilde{\alpha}_1 = \left( -1, 1, 0\right) \,\,\, , \tilde{\alpha}_2 =
\left( 0, -1 , 1\right) \,\,\, , \,\,\, \tilde{\alpha}_3 = \left( 1
,1, 0\right) ,
\end{equation}
since this basis generates the same lattice. Now, the projection 
on the 1-2 plane yields vectors $\left( -1 ,0\right)$, $\left( 0,
-1\right)$ and $\left( 1, 1\right)$ and there is no non degenerate
coordinate transformation such that one of these vectors becomes
orthogonal to the two others. So, our previous argument for the
equivalence to an SO(4)$\times$SU(2) lattice seems to fail.   
However, the systems (\ref{eq:so6}) and (\ref{eq:so6t}) are related by
a coordinate transformation
\begin{equation}\label{eq:cocha}
\tilde{x}^1 = x^3 \,\,\, , \,\,\, \tilde{x}^2 = x^2 \,\,\, , \,\,\,
\tilde{x}^3 = x^1 .
\end{equation}
The metric component $G_{13}$  is projected out by the orbifold.
The coordinate transformation (\ref{eq:cocha}) does not generate such
a component from a vanishing one and hence does not induce a projected
metric deformation. (Indeed, the Euclidean metric is invariant under
(\ref{eq:cocha}).) So, we can safely go back to the choice
(\ref{eq:so6}) and perform the previous deformations to connect to the
SO(4)$\times$SU(2) lattices. The conclusion from this short discussion
is that the basis of a lattice can be replaced by another basis of the
same lattice. This should be always the case as long as the orbifold
acts as a lattice automorphism. (This statement seems somewhat
trivial. We mentioned it, even so, since it will be important for later
applications.)     

In summary, our examples illustrate that whenever the orbifold action
is such that in a Cartesian basis there are off diagonal metric
moduli, the compactification lattice can be continuously deformed into
a lattice where one of the basis vectors is orthogonal to the rest of
the lattice. This is the case when the $\theta_1$ as well as
the $\theta_2$ action on a two dimensional subspace of ${\mathbb R}^3$
equals $\pm\id_2$.   

\section{\boldmath{${\mathbb Z}_2 \times {\mathbb Z}_2$} Orbifolds of
  \boldmath{$T^6$}} 

Our ultimate goal is it to construct phenomenologically interesting
string models. A prominent route to take is to compactify heterotic
string theory on a six dimensional orbifold. Such models easily
lead to $N=1$ supersymmetric theories with non Abelian gauge groups
and chiral matter in four dimensions. The requirement of having
exactly $N=1$ unbroken supersymmetry in four dimensions imposes
constraints on the orbifold action. In the present paper, we are
interested in one of the simplest orbifold groups {\it viz.} ${\mathbb
  Z}_2 \times {\mathbb Z}_2$. In order to obtain $N=1$ supersymmetry
in four dimensions we have to specify the action of the two ${\mathbb
  Z}_2 \times {\mathbb Z}_2$ generators, $\theta_1$ and $\theta_2$, on
a Cartesian coordinate system as follows
\begin{equation}\label{eq:6dorb}
\theta_1 = diag\left( -1, -1, -1, -1, 1,1\right) \,\,\, , \,\,\,
\theta_2 = diag\left( 1,1, -1, -1, -1, -1\right) .
\end{equation}
This choice is unique up to relabelling the coordinates.  

Next, we would like to classify all possible compactification
lattices. From the action of the orbifold elements (\ref{eq:6dorb})
and the conclusions of the previous section we see that whenever a
lattice extends over more then three directions we can continuously
deform it to a lattice consisting of two orthogonal
factors. Therefore, any six dimensional lattice can be deformed into a
product of mutually orthogonal three dimensional lattices and we are
left with the task of classifying three dimensional lattices. 
 
Before doing so, let us illustrate the above statement at the example
of an SO(12) compactification lattice. The basis vectors of the
compactification lattice are the SO(12) simple roots
\begin{eqnarray}
\alpha_1 & = & \left( 1 , -1 , 0,0,0,0\right) , \nonumber \\
\alpha_2 & = & \left( 0, 1, -1, 0,0,0\right) , \nonumber \\
\alpha_3 & = & \left( 0 ,0 , 1, -1,0, 0\right) , \nonumber \\
\alpha_4 & = & \left( 0, 0, 0, 1,-1,0\right) , \nonumber \\
\alpha_5 & = & \left( 0,0,0,0,1, -1\right) , \nonumber \\
\alpha_6 & = & \left( 0,0,0,0,1, 1\right) .\label{eq:so12roots}
\end{eqnarray}
The geometric moduli correspond to changes of the coordinate basis in
the 1-2, 3-4 and 5-6 plane while keeping the metric fixed, and, as
discussed in the end of the previous section, changing the basis of a
lattice. Changing $x^2 \to x^2 + x^1$ we can `decouple' $\alpha_1$,
and (after rescaling $x^1 \to \sqrt{2} x^1$) obtain an
SU(2)$\times$SO(10) lattice. For the SO(10) lattice we replace the
basis $\alpha_2, \ldots , \alpha_6$ by an equivalent one and obtain
\begin{eqnarray}
e_1  & = & \left( \sqrt{2},0,0,0,0,0\right) , \nonumber \\
\tilde{\alpha}_2 & = & \left( 0, 1, -1,0,0,0\right) , \nonumber \\
\tilde{\alpha}_3 & = & \left( 0, 1 , 1,0,0,0\right) , \nonumber \\
\tilde{\alpha}_4 & = & \left( 0, -1, 0 ,1, 0,0\right) , \nonumber \\
\tilde{\alpha}_5 & = & \left( 0,0,0,-1, 1,0\right) , \nonumber \\  
\tilde{\alpha}_6 & = & \left( 0,0,0,0,-1,1\right) .
\end{eqnarray}
Now, we perform another  allowed metric deformation by replacing $x^5
\to x^5 + x^6$ and $x^6 \to \sqrt{2}x^6$. Then $\tilde{\alpha}_2,
\ldots \tilde{\alpha}_5$ generate an SO(8) root lattice whereas
$\tilde{\alpha}_6$ generates another orthogonal SU(2) lattice. For the
SO(8) lattice we pick again an equivalent set of basis vectors
\begin{eqnarray}
e_1  & = & \left( \sqrt{2},0,0,0,0,0\right) , \nonumber \\
\hat{\alpha}_2 & = & \left( 0, 1,0,0, -1,0\right) , \nonumber \\
\hat{\alpha}_3 & = & \left( 0, 1 ,0,0, 1,0\right) , \nonumber \\
\hat{\alpha}_4 & = & \left( 0, -1, 1 ,0, 0,0\right) , \nonumber \\
\hat{\alpha}_5 & = & \left( 0,0,-1, 1,0,0\right) , \nonumber \\  
e_6 & = & \left( 0,0,0,0,0,\sqrt{2}\right) .
\end{eqnarray}
Finally, we deform geometric moduli in the 3-4 plane by replacing $x^3
\to x^3 + x^4$ and $x^4 \to \sqrt{2} x^4$ we obtain as a
compactification lattice the root lattice of\footnote{From now on we drop
  our previous differentiation of SO(6) and SU(4) lattices.}
SU(4)$\times$SU(2)$^3$  
\begin{eqnarray}
e_1  & = & \left( \sqrt{2},0,0,0,0,0\right) , \nonumber \\
\hat{\alpha}_2 & = & \left( 0, 1,0,0, -1,0\right) , \nonumber \\
\hat{\alpha}_3 & = & \left( 0, 1 ,0,0, 1,0\right) , \nonumber \\
\hat{\alpha}_4 & = & \left( 0, -1, 1 ,0, 0,0\right) , \nonumber \\
e_5 & = & \left( 0,0,0, \sqrt{2},0,0\right) , \nonumber \\  
e_6 & = & \left( 0,0,0,0,0,\sqrt{2}\right) .
\end{eqnarray}
The SU(4) lattice, generated by $\hat{\alpha}_2$, $\hat{\alpha}_3$,
$\hat{\alpha}_4$, is extended along the 2, 3 and 5 direction, and there
are not enough geometric moduli to decompose it further ($G_{23}$,
$G_{25}$ and $G_{35}$ are projected out by the orbifold action). 
The SO(12) example illustrates our previous statement that three
dimensional factorisable and non-factorisable lattices appear as
building blocks for $T^6/{\mathbb Z}_2 \times {\mathbb Z}_2$ orbifolds
when the orbifold action is given by (\ref{eq:6dorb}) such that $N=1$
supersymmetry is unbroken. In terms of non-factorisable lattices the
building blocks are thus given by three, two and one dimensional
lattices. We are going to classify these in the following.

A non-factorisable three dimensional lattice cannot be decomposed
further if the orbifold action on the three dimensional subspace is
given by (up to permuting $\theta_1$, $\theta_2$ and $\theta_3 =
\theta_1 \theta_2$)
\begin{equation}
\theta_1=(-1,-1,1), \qquad \theta_2=(1,-1,-1), \qquad
\theta_3=(-1,1,-1) .
\label{su(4)-twist}
\end{equation}
We use the basis of simple roots (\ref{eq:so6}), and twists
$\theta_i$ (\ref{su(4)-twist}). 
(We obtain the same result when we use the basis (\ref{eq:su4}).)

 The twists $\theta_i$ are represented by Weyl reflections and outer
automorphism.
For example, $\theta_1 $ is a product of the outer automorphism
swapping $\alpha_1$ with $\alpha_3$ and a total ${\mathbb Z}_2$
rotation,
$\alpha_i \rightarrow - \alpha_i$, while the twist $\theta_2$ is a sum of
two Weyl reflections at $\alpha_1$ and $\alpha_3$.
Thus, the twists $\theta_i$ transform the SU(4) simple roots as
\begin{eqnarray}
 \label{su4-twist}
& &
\theta_1 \alpha_1 = -\alpha_3, \qquad
\theta_1 \alpha_2 = -\alpha_2, \qquad
\theta_1 \alpha_3 = -\alpha_1,  \nonumber \\
 & &
\theta_2 \alpha_1 = -\alpha_1, \qquad
\theta_2 \alpha_2 = \sum_{i=1}^3 \alpha_i, \qquad
\theta_2 \alpha_3 = -\alpha_3,  \\
 & &
\theta_3 \alpha_1 = \alpha_3, \qquad
\theta_3 \alpha_2 = -\sum_{i=1}^3 \alpha_i, \qquad
\theta_3 \alpha_3 = \alpha_1.
\nonumber
\end{eqnarray}
The sublattices $(1 - \theta_i)\Lambda_{SU(4)}$ and the
normal lattices $N_{\theta_i}$ are spanned by
\begin{equation}
\begin{array}{ll}
 (1 -\theta_1) \Lambda_{SU(4)}:\{ (\alpha_1 + \alpha_3),2\alpha_2\}, &\qquad
N_{\theta_1}: \{(\alpha_1 + \alpha_3),\alpha_2\}, \\
 (1 -\theta_2) \Lambda_{SU(4)}:\{ (\alpha_1 +\alpha_3),2\alpha_1 \},
& \qquad N_{\theta_2}: \{ \alpha_1,\alpha_3 \},  \\
 (1-\theta_3)\Lambda_{SU(4)}:\{ 2(\alpha_2 + \alpha_3), (\alpha_1-\alpha_3) \},
& \qquad N_{\theta_3}: \{ (\alpha_2 + \alpha_3), (\alpha_1-\alpha_3) \}.
\end{array}
\end{equation}
Thus, each of the $T_1$, $T_2$ and $T_3$ sectors contains two fixed tori,
\begin{equation}
T_1: (\theta_1,m^{(1)} \alpha_2), \qquad
T_2: (\theta_2,m^{(2)} \alpha_1), \qquad
T_3: (\theta_3,m^{(3)}(\alpha_1+\alpha_2)),
\end{equation}
where $m^{(a)}=0,1$ for $a=1,2,3$.

A non-factorisable two dimensional lattice cannot be decomposed
further by deforming geometric moduli if the orbifold acts on the two
dimensional subspace as
\begin{equation}
\theta_1 = diag\left( -1,1\right) \,\,\, , \,\,\, \theta = - \id_2 .
\end{equation}
First, we take the following SU(3) lattice as a compactification
lattice:
\begin{equation}\label{eq:su3}
\alpha_1 = \left( \sqrt{2}, 0 \right) \,\,\, , \,\,\, \alpha_2 =
\left( -\frac{1}{\sqrt{2}}, \sqrt{\frac{3}{2}}\right) .
\end{equation}
On this root lattice $\theta_1$ acts as a Weyl reflection at
$\alpha_1$, and $\theta_2$ as multiplying all roots with a minus sign 
\begin{eqnarray}
\label{su3-twist}
& &
\theta_1 \alpha_1 = - \alpha_1, \qquad
\theta_1 \alpha_2 = \alpha_1 + \alpha_2, \nonumber \\
 & &
\theta_2 \alpha_1 = - \alpha_1, \qquad
\theta_2 \alpha_2 = - \alpha_2, \\
  & &
\theta_3 \alpha_1 = \alpha_1, \qquad
\theta_3 \alpha_2 = -\alpha_1 - \alpha_2 . \nonumber
\end{eqnarray}
The sublattices $(1 - \theta_i) \Lambda_{SU(3)}$ and the normal
lattices $N_{\theta_i}$ are obtained as
\begin{equation}
\begin{array}{ll}
 (1 -\theta_1) \Lambda_{SU(3)}:\{ \alpha_1 \}, &\qquad
N_{\theta_1}: \{\alpha_1\}, \\
 (1 -\theta_2) \Lambda_{SU(3)}:\{ 2\alpha_1,2\alpha_2 \},
& \qquad N_{\theta_2}: \{ \alpha_1,\alpha_2 \},  \\
 (1-\theta_3)\Lambda_{SU(3)}:\{ \alpha_1 +2\alpha_2 \},
& \qquad N_{\theta_3}: \{ \alpha_1 +2\alpha_2  \}.
\end{array}
\end{equation}
The $T_1$, $T_2$ and $T_3$ sectors have one, four and one fixed tori,
\begin{equation}
T_1:~(\theta_1,0), \qquad T_2:(\theta_2,m^{(2)}_1\alpha_1 +
m^{(2)}_2\alpha_2), \qquad T_3:(\theta_3,0),
\end{equation}
where $m^{(2)}_1,m^{(2)}_2=0,1$.

Now, we are going to argue that the SU(3) lattice (\ref{eq:su3}) can
be deformed into an SO(4) lattice. At first sight this is not obvious
since the only allowed deformations are separate rescalings of the two
directions. However, the same SU(3) root lattice is generated by the
alternative basis
\begin{equation} \label{eq:su3p}
\alpha_1 = \left( \frac{1}{\sqrt{2}}, \sqrt{\frac{3}{2}}\right) \,\,\,
, \,\,\, \alpha_2 = \left( \frac{1}{\sqrt{2}},
-\sqrt{\frac{3}{2}}\right) .
\end{equation}
Now, $\theta_1$ acts as an outer automorphism on the simple roots
 \begin{eqnarray}
\label{so4-twist}
 & & \theta_1 \alpha_1 = \alpha_2, \qquad \theta_1 \alpha_2 =
 \alpha_1,
\nonumber \\
 & & \theta_2 \alpha_1 = -\alpha_1, \qquad \theta_2 \alpha_2 =
    -\alpha_2,   \\
 & & \theta_3 \alpha_1 = -\alpha_2, \qquad \theta_3 \alpha_2 =
    -\alpha_1. \nonumber
\end{eqnarray}
The system (\ref{eq:su3p}) can be geometrically deformed into
an SO(4) root lattice. By rescaling the two directions one can bring
it to the form
\begin{equation}\label{eq:so4}
\alpha_1 = \left( 1 , 1\right) \,\,\, , \,\,\, \alpha_2 = \left( 1 ,
-1\right) .
\end{equation}
These vectors span an SO(4) root lattice, which we distinguish from an
SU(2)$^2$ lattice by noticing that geometric moduli do not allow
for separate changes in the length of the two roots.  Below, we will
use the SU(3) lattice (\ref{eq:su3}) for the classification and not
its equivalent versions (\ref{eq:su3p}) or (\ref{eq:so4}). 

In contrast to the SO(4) lattice (\ref{eq:so4}), an SU(2)$\times$SU(2)
lattice will be given by  
\begin{equation}
\alpha_1 = \left( \sqrt{2}, 0 \right) \,\,\, ,\,\,\, \alpha_2 = \left(
0,\sqrt{2}\right) .
\end{equation}
This factorises into two one dimensional lattices. So, finally we
consider a one dimensional lattice spanned by $e_1$. The three
twists $\theta_i$ $(i=1,2,3)$ are defined as 
\begin{equation}
\label{su2-twist}
\theta_1 e_1=-e_1, \qquad \theta_2 e_1 = - e_1,
\qquad \theta_3 e_1 =e_1.
\end{equation}
The sublattices $(1 - \theta_i) \Lambda_{SU(2)}$ and the normal
lattices $N_{\theta_i}$ are obtained as
\begin{equation}
\begin{array}{ll}
 (1 -\theta_1) \Lambda_{SU(2)}:\{ 2e_1\}, &\qquad
N_{\theta_1}: \{e_1 \}, \\
 (1 -\theta_2) \Lambda_{SU(2)}:\{ 2e_1 \},
& \qquad N_{\theta_2}: \{ e_1 \}.
\end{array}
\end{equation}
Both $T_1$ and $T_2$ sectors have two fixed tori,
\begin{equation}
T_1:~(\theta_1,m^{(1)}e_1), \qquad T_2:(\theta_2,m^{(2)}_1e_1 ),
\end{equation}
where $m^{(1)},m^{(2)}=0,1$.

In summary, the building blocks of supersymmetric $T^6/{\mathbb Z}_2
\times {\mathbb Z}_2$ are the three dimensional SU(4) root lattice,
the two dimensional SU(3) root lattice (\ref{eq:su3}) and the one
dimensional SU(2) root lattice. Our results, including also the fixed
point structure, are listed in Table \ref{tab:build}.
\begin{table}
\begin{center}
\begin{tabular}{|c||c|c|c|c||c|c|c|} \hline
%\footnotesize
orbifold & $\alpha_i$ & $\theta_1$ & $\theta_2$ & $\theta_3$
        & $T_1(\theta_1,V_1)$ & $T_2(\theta_2,V_2)$ & $T_3(\theta_3,V_3)$
\\  \hline \hline
$SU(4)$ & $\alpha_1$ & $ -\alpha_3$
        & $ -\alpha_1$
        & $ \alpha_3$   & & & \\
        & $\alpha_2$ & $ -\alpha_2$
        & $ \sum_i^3 \alpha_i$
        & $ - \sum_i^3 \alpha_i$
        &  $m^{(1)} \alpha_2$
        & $m^{(2)} \alpha_1$
        & $m^{(3)}(\alpha_1+\alpha_2)$   \\
        & $\alpha_3$ & $ -\alpha_1$
        & $ -\alpha_3$
        & $ \alpha_1$ & & & \\ \hline
$ SU(3)$ & $\alpha_1$ &  $ -\alpha_1 $
         & $ -\alpha_1$
         & $ \alpha_1$
         & $0$
         & $m^{(2)}_1\alpha_1 + m^{(2)}_2\alpha_2 $
         & $0$ \\
         & $\alpha_2$ &$\sum_i^2 \alpha_i$
         & $ -\alpha_2$
         & $ -\sum_i^2 \alpha_i$ & & & \\ \hline
$SU(2)$ & $e_1$ & $-e_1$ & $-e_1$ & $e_1$ & $ m^{(1)}e_1$ 
& $m^{(2)}_1e_1$ & \\ \hline 
\end{tabular}
\caption{Building blocks for ${\mathbb Z}_2 \times {\mathbb Z}_2$
  orbifolds} 
\label{tab:build}
\end{center}
\end{table}

From these building blocks we can construct altogether eight classes of
$T^6/ {\mathbb Z}_2 \times {\mathbb Z}_2$ orbifolds which are listed
in the Appendix.

\section{Chiral Spectra for Standard Embedding \label{sec:standard}}

In ${\mathbb Z}_2 \times {\mathbb Z}_2$ orbifolds a chiral spectrum
originates from twisted sectors. Hence, a modification of the fixed
point structure affects the chiral matter content of the model. 

For example, if the orbifold action is standard embedded into the E$_8
\times$E$_8$ gauge group of the ten dimensional heterotic string the
unbroken gauge group in four dimensions contains an E$_6$ factor
\cite{Font:1988mk}. The untwisted sector gives rise to three chiral
multiplets in the {\bf 27} and three chiral multiplets in the
{\boldmath{$\overline{27}$}} of E$_6$, whereas the twisted sectors
provide different numbers of {\bf 27}'s and
{\boldmath{$\overline{27}$}}'s. 

As in \cite{Faraggi:2006bs} we could obtain the chiral spectrum by
explicit construction. There is, however, a shortcut employing 
modular invariance of the partition function (see e.g.\ the third
reference in \cite{IMNQ} and \cite{Erler:1992ki}). In
\cite{Faraggi:2006bs} this method is briefly described in the context
of non factorisable ${\mathbb Z}_2 \times {\mathbb Z}_2$ orbifolds. In
the following we repeat that discussion giving also more details. In
the end, we obtain a closed formula for the number of generations 
({\bf 27}'s) and anti-generations ({\boldmath{$\overline{27}$}}'s) for
the standard embedded orbifold. (For embeddings other than the standard
embedding and in the presence of Wilson lines the same method works,
the E$_8 \times$E$_8$ sector of the partition function has to be
considered explicitly, though.)

We write the partition function as 
\begin{equation}
Z\left( \tau, \overline{\tau}\right) = \sum_{i,j} \Delta_{i,j} Z_{i,j} ,
\end{equation}
where $i,j$ label the elements of the orbifold group, for example
\begin{equation}
Z_{\theta_1 ,\theta_2} = \frac{1}{4} \mathop{\mbox{Tr}}_{\mbox{\tiny $\theta_1$
    twisted }}\theta_2 q^{L_0} \overline{q}^{\overline{L_0}},
\end{equation}
with $q = \exp \left( 2 \pi i \tau\right)$. The factors $\Delta_{i,j}$
are fixed by modular invariance. In the untwisted sector they are the
same as in the uncompactified case (e.g.\ one for bosons). The number
of generations can be determined by identifying the contributions from
massless chiral fermions to $Z$. 

In the $\theta_1$ twisted sector, for example, terms with an identity
or $\theta_1$ insertion in the trace have identical contributions from
GSO even chiral and anti-chiral fermions. For terms with a $\theta_2$
or $\theta_1 \theta_2$ insertion, GSO even chiral and anti-chiral
fermions contribute with opposite signs.\footnote{The corresponding
  $\Delta$'s 
seem not to be fixed by modular invariance since they multiply
zero. This will, however, not be important for the discussion
below. Changing their sign is called discrete torsion which for
${\mathbb Z}_2 \times {\mathbb Z}_2$ orbifolds just swaps the number
of families with the number of anti-families
\cite{Font:1988mk,Vafa:1994rv}.} 
We find for the number of chiral fermions, $M$, and anti-chiral fermions,
$N$, from the $\theta_1$ twisted sector 
\begin{eqnarray}
M & = & \tilde{\Delta}_{\theta_1, 1} + \tilde{\Delta}_{\theta_1, \theta_1} +
\tilde{\Delta}_{\theta_1 , \theta_2} + \tilde{\Delta}_{\theta_1,
  \theta_1 \theta_2} , 
\\
N & = &  \tilde{\Delta}_{\theta_1, 1} + \tilde{\Delta}_{\theta_1, \theta_1} -
\tilde{\Delta}_{\theta_1 , \theta_2} - \tilde{\Delta}_{\theta_1,
  \theta_1 \theta_2} ,
\end{eqnarray}
where the tilde indicates absorption of numerical factors coming from
the trace. For the factorisable case we know that there are 16
generations and zero anti-generations. Therefore we must have
\begin{equation}
\tilde{\Delta}_{\theta_1, 1} + \tilde{\Delta}_{\theta_1, \theta_1} 
=\tilde{\Delta}_{\theta_1 , \theta_2} +\tilde{\Delta}_{\theta_1,
  \theta_1 \theta_2} = 8
\end{equation}
in the factorisable case. 

What we will do now is to identify dependencies on the compactification
lattice  in the $\tilde{\Delta}$'s and in turn fix their value by our
knowledge of the factorisable case. 
For example, in the contribution from untwisted states with a
$\theta_1$ insertion in the trace there will be a sum over winding and
momenta corresponding to the two compact directions which are not
affected by $\theta_1$. The same happens in the $\theta_1$ twisted
sector if an identity or $\theta_1$ is inserted into the trace. A
modular transformation takes the untwisted sector sums to the twisted
sector ones which can be seen by Poisson resumming winding and
momenta. 

For non factorisable compactifications there are two types of two-tori
appearing. One comes from the $\theta_1$ projection of the
compactification lattice. (This is not necessarily a sublattice.) The
other one is spanned by an invariant sublattice of the
compactification lattice. For the contribution from the untwisted
sector with $\theta_1$ insertion, windings are summed over the
invariant lattice whereas momenta are summed over the invariant dual
lattice (only these contribute to the trace). A subtlety for non
factorisable compactifications is that the invariant dual lattice is
not the dual of the invariant lattice but the dual of the projected
lattice as can be seen from the defining property
\begin{equation}
\delta_{ij} = \left( e_i ^\star , e_j\right) = \left( 
  \frac{1 +\theta_1 ^T}{2} e^\star _i \, ,\, 
e_j\right) =  \left( e^\star _i \, ,\, \frac{1 +\theta_1}{2}
e_j\right) .
\end{equation}
For the twisted sector windings are on the projected lattice (strings
do have to close only up to $\theta_1$ identifications). Momenta are
quantised w.r.t.\ to the dual of the invariant sublattice which spans the
fixed torus. 

After Poisson resumming the untwisted sector the erstwhile momentum
sum becomes a winding sum over the projected lattice whereas the
erstwhile winding sum becomes a momentum sum over the dual of the
invariant lattice. This is consistent with modular invariance. The
Poisson resummations provide also a numerical factor of
the ratio\footnote{The volume of a dual lattice is inverse to the
  volume of the lattice.} between the volume of the projected lattice
  (vol$\left( 
\Lambda^\perp\right)$) and the volume of the invariant lattice 
(vol$\left(\Lambda^{\mbox{\tiny inv}}\right)$). Hence we conclude
\begin{equation}
\tilde{\Delta}_{\theta_1, 1} + \tilde{\Delta}_{\theta_1, \theta_1} = 8
\frac{ \mbox{vol}  \left(
  \Lambda^\perp\right)}{\mbox{vol}\left(\Lambda^{\mbox{\tiny
      inv}}\right)} ,
\end{equation}
where we used that for the known factorisable case the ratio of the
volumes is one.

In the traces over the $\theta_1$ twisted sector with a $\theta_2$ or
$\theta_1\theta_2$ insertion there will be no sum over windings or
momenta since these quantum numbers are not invariant under the
inserted operator if they differ from zero. There is, however, also a
dependence on the compactification lattice. Only if the  position of a
$\theta_1$ 
twisted state is invariant under the inserted operator that state will
contribute to the trace. Calling the number of points, which are
invariant under $\theta_i$ and $\theta_j$, $\chi\left( \theta_i,
\theta_j\right)$  and comparing to the factorisable case (in which
these numbers are 64) we find
\begin{equation}
\tilde{\Delta}_{\theta_1,\theta_2} +
  \tilde{\Delta}_{\theta_1,\theta_1\theta_2} = \frac{\chi\left( \theta_1,
  \theta_2\right)}{8} ,
\end{equation}
where we used the obvious relation $\chi \left( \theta_1,
\theta_2\right) = \chi \left( \theta_1, \theta_1 \theta_2\right)$.

Collecting everything we obtain the following relations for the number
of families $M$ and anti-families $N$ from imposing modular invariance
\begin{eqnarray}
M & = &  8
\frac{ \mbox{vol}  \left(
  \Lambda^\perp\right)}{\mbox{vol}\left(\Lambda^{\mbox{\tiny
      inv}}\right)} + \frac{\chi\left( \theta_1,
  \theta_2\right)}{8} , \\
N & = & 8\frac{ \mbox{vol}  \left(
  \Lambda^\perp\right)}{\mbox{vol}\left(\Lambda^{\mbox{\tiny
      inv}}\right)} - \frac{\chi\left( \theta_1,
  \theta_2\right)}{8} .
\end{eqnarray}
We added the resulting numbers for families and anti-families to the
list of models in the Appendix.

\section{Further Phenomenological Aspects}

\subsection{Wilson Line}

Twisted sector states corresponding to the same orbifold group element
but to a different fixed point or torus form identical four
dimensional spectra.
Wilson lines can lift that degeneracy. Thus, they
provide an important tool for the reduction of the number of families
\cite{IMNQ, Kobayashi:1990mi, Kobayashi:1991rp}.
Here, we study discrete Wilson lines possible in
3D SU(4), 2D SU(3) and 1D SU(2)
${\mathbb Z}_2 \times {\mathbb Z}_2$ orbifolds.

A Wilson line can be viewed as a non trivial embedding of a shift by
a lattice vector into the gauge group. In the bosonic formulation of
the heterotic E$_8 \times$E$_8$ theory, there are 16 left moving
bosons compactified on an E$_8 \times$E$_8$ root lattice.
Using that language, we embed the space group element $(\theta, V)$ as
a shift $(V_\theta + W_V)$ into the 16 left moving directions.

The shift vector $V$ of  6D compact space is embedded as the shift
$W_V$ into the E$_8 \times$E$_8$ lattice, and $W_V$ is called Wilson
line along the cycle $V$.
Different fixed points (tori) in the same twisted sector
are fixed under the action of space group elements with different
shifts in the 6D space. 
Thus, Wilson lines can lift the degeneracy of the spectrum
within the same twisted sector.
Since
$V$ and $\theta V$ are identical on the orbifold, the corresponding
Wilson lines should be equivalent, i.e.\ differ at most by an
$\Lambda_{E_8 \times E_8}$ lattice vector. 

As an example we consider the 3D SU(4) ${\mathbb Z}_2 \times {\mathbb
  Z}_2$ 
orbifold with the twist (\ref{su4-twist}).
The twist $\theta_1$ requires that
$W_{\alpha_1} + W_{\alpha_3}$ and $2W_{\alpha_2}$
should be on $\Lambda_{E_8 \times E_8}$,
where $W_{\alpha_i}$ denotes the Wilson line along
$\alpha_i$.
Furthermore, the twist $\theta_2$ requires
$2W_{\alpha_1}$ and $2W_{\alpha_3}$ to be on
$\Lambda_{E_8 \times E_8}$.
As a result, we obtain the condition for Wilson lines as
\begin{equation}
2W_{\alpha_1} = 2W_{\alpha_2} = 0, \qquad
W_{\alpha_1} = W_{\alpha_3},
\end{equation}
up to $\Lambda_{E_8 \times E_8}$. 

Similarly, we can study conditions on Wilson lines for
other building blocks, that is,
SU(3) and SU(2) ${\mathbb Z}_2 \times {\mathbb Z}_2$
orbifolds. Results are shown in Table \ref{tab:wilcond}.
In addition to these geometrical
consistency conditions one has to impose modular invariance conditions
\cite{IMNQ} when including Wilson lines in the full heterotic
construction. 

\begin{table}
\begin{center}
\begin{tabular}{|c|c|c|}\hline
orbifold & twists & conditions for Wilson lines (up to
$\Lambda_{E_8 \times E_8}$) \\ \hline \hline
$SU(4)$ & Eq.~(\ref{su4-twist}) &
$2W_{\alpha_1} = 2W_{\alpha_2} = 0,$~~~$W_{\alpha_1} = W_{\alpha_3}$
\\ \hline
$SU(3)$ & Eq.~(\ref{su3-twist}) &
$W_{\alpha_1} = 0,$~~~$2W_{\alpha_2} = 0$
\\ \hline
$SU(2)$ & Eq.~(\ref{su2-twist}) &
$2W_{\alpha_1} = 0$
\\ \hline
\end{tabular}
\caption{Conditions for Wilson lines. \label{tab:wilcond}}
\end{center}
\end{table}

\subsection{Yukawa Couplings}

Here we consider Yukawa couplings in non-factorisable
${\mathbb Z}_2 \times {\mathbb Z}_2$ orbifold models.
We focus on the space group selection rule
for allowed Yukawa couplings, because selection rules for
other parts are the same as in conventional
factorisable models.
In particular, we study the 3-point coupling of three twisted states
corresponding to the space group elements,
$(\theta_1,V_1)$, $(\theta_2,V_2)$ and $(\theta_3,V_3)$.
Their coupling is allowed if the product of three
space group elements is equal to identity.
Note that the space group element $(\theta_i,V_i)$ is
equivalent to $(\theta_i,V_i + (1 - \theta_i)\Lambda)$,
up to conjugacy class.
Therefore,
the condition for allowed coupling is obtained as
\cite{Dixon:1986qv,Kobayashi:1991rp}
\begin{equation}
(\theta_1,V_1)(\theta_2,V_2)(\theta_3,V_3) =(1,0),
\end{equation}
up to $(1 - \theta_i)\Lambda$.

As an example, let us consider the SU(4) ${\mathbb Z}_2 \times
{\mathbb Z}_2$ orbifold.
Then, we study the 3-point coupling corresponding to
three twisted states, $(\theta_1,m^{(1)} \alpha_2)$,
$(\theta_2,m^{(2)} \alpha_1)$ and
$(\theta_3,m^{(3)}(\alpha_1+\alpha_2))$.
Note that $\cup_i (1-\theta_i)\Lambda_{SU(4)}$ is spanned by
$2\alpha_1$ and $2\alpha_2$ as well as $\alpha_ 1 -\alpha_3$.
Thus, this coupling is allowed when
\begin{equation}
m^{(1)} + m^{(3)} = {\rm ~~even},\qquad
m^{(2)} + m^{(3)} = {\rm ~~even}.
\end{equation}

For the other orbifolds, we can obtain conditions on allowed couplings
in a similar way.
Results are shown in Table \ref{tab:yuk}.
The fifth column shows conditions for allowed $T_1T_2T_3$ couplings.
Note that only $T_1T_2T_3$ type of couplings are allowed, while
the other types, e.g. $T_i T_iT_j$ ($i\neq j$) and
$T_i T_i T_i$ are forbidden by the point group selection rule.
Only diagonal couplings are allowed on 3D SU(4) and 1D SU(2)
${\mathbb Z}_2 \times {\mathbb Z}_2$ orbifolds, that is, when we chose
two states, 
the other state to be allowed to couple is uniquely fixed.
However, off-diagonal couplings are allowed on 2D SU(3)/${\mathbb Z}_2
\times {\mathbb Z}_2$ orbifolds. 
On the 2D SU(3)/${\mathbb Z}_2 \times {\mathbb Z}_2$ orbifold,
both states corresponding to $(\theta_2,m^{(2)}_1 \alpha_1)$
for $m^{(2)}_1=0,1$ in the $T_2$ sector can couple with
the $T_1$ and $T_3$ sectors corresponding to
$(\theta_1,0)$ and $(\theta_3,0)$.

Here, we note an (exclusion) relation between the possibility of having
off-diagonal couplings and discrete Wilson lines.
For example, the 2D SU(3) ${\mathbb Z}_2 \times {\mathbb Z}_2$
orbifold does not allow 
non-trivial Wilson lines along $\alpha_1$, i.e.\ $W_{\alpha_1}=0$ up
to $\Lambda_{E_8 \times E_8}$. 
Unfortunately, it is not clear how discrete Wilson lines contribute
to magnitudes and CP phases of Yukawa couplings.
However, such ignorance does not affect two types of
couplings, $(\theta_1,0)(\theta_2,0)(\theta_3,0)$ and
$(\theta_1,0)(\theta_2,\alpha_1)(\theta_3,0)$,
because there is no non-trivial discrete Wilson line $A_{\alpha_1}=0$.
Thus, when off-diagonal couplings are allowed,
there is no non-trivial discrete Wilson lines to distinguish them.
This aspect has been found, previously,  in ${\mathbb Z}_N$ orbifold
models \cite{Kobayashi:2003gf}.

\begin{table}
\begin{center}
\begin{tabular}{|c|c|c|c|c|} \hline
orbifold & $T_1$ & $T_2$ & $T_3$ & Conditions  \\ \hline \hline
$SU(4)$ & $(\theta_1,m^{(1)} \alpha_2)$
        & $(\theta_2,m^{(2)} \alpha_1)$
        & $(\theta_3,m^{(3)}(\alpha_1+\alpha_2))$
        & $m^{(1)} + m^{(3)} = $~~even \\
        & & & & $m^{(2)} + m^{(3)} = $~~even
        \\ \hline
$SU(3)$ & $(\theta_1,0)$
        & $(\theta_2,m^{(2)}_1\alpha_1 + m^{(2)}_2\alpha_2) $
        & $(\theta_3,0)$
        & $m^{(2)}_2=0$ \\ \hline
$SU(2)$ & $(\theta_1,m^{(1)}e_1)$
        & $(\theta_2,m^{(2)}_1e_1)$ &
        & $m^{(1)} + m^{(2)} = $~~even \\ \hline
\end{tabular}
\caption{Conditions for allowed $T_1T_2T_3$ couplings. \label{tab:yuk}}
\end{center}
\end{table}

Magnitudes of allowed Yukawa couplings can be computed
by the usual method \cite{Hamidi:1986vh,Dixon:1986qv,
Burwick:1990tu,Kobayashi:2003vi}.
In these models, all of allowed Yukawa couplings are of $O(1)$
w.r.t.\ the string scale up to field redefinitions needed for
obtaining canonical K\"ahler potentials.

\section{Conclusion and Discussion}

In the present paper, we classified ${\mathbb Z}_2 \times {\mathbb
  Z}_2$ orbifolds of non-factorisable six-tori. We restricted our
  attention to orbifolds that lead, if taken as the compact space for
  heterotic strings, to $N=1$ supersymmetry in four dimensions. We
  found that all topologically inequivalent orbifolds can be composed
  from the following building blocks:
3D SU(4), 2D SU(3) and 1D
SU(2) ${\mathbb Z}_2 \times {\mathbb Z}_2$ orbifolds.
Using them, we have classified eight 6D ${\mathbb Z}_2 \times {\mathbb
  Z}_2$ orbifolds, 
including the conventional factorisable one.
In the factorisable ${\mathbb Z}_2 \times {\mathbb Z}_2$ orbifold models,
each of the three twisted sectors has 16 fixed tori, whereas
for non-factorisable orbifold models these numbers are smaller.
In the model with the minimal numbers,
each of three twisted sectors has four fixed tori.
Thus, these models have a variety of generation numbers.
For example, in the standard embedding, the smallest number 
of net generations among these eight classes is equal to six, 
while the largest number is 48, which is obtained in the 
conventional factorisable model.
We have also studied discrete Wilson lines and
selection rules for allowed Yukawa couplings.
These features are different from the conventional factorisable
${\mathbb Z}_2 \times {\mathbb Z}_2$ orbifold models.

The non-factorisable models allow off-diagonal couplings, while
the factorisable one allows only diagonal couplings.
However, it seems difficult to realise quark/lepton
masses and mixing angles by use of only 3-point couplings
in non-factorisable ${\mathbb Z}_2 \times {\mathbb Z}_2$ orbifold models with
the minimal number of Higgs fields.

In order to obtain realistic fermion mass matrices,
it would be interesting to introduce more Higgs fields and/or
use higher dimensional operators.
The selection rules for generic $n$-point couplings can be
obtained by extending our analysis of the 3-point couplings.
Furthermore, it is important to study
which type of non-Abelian flavour symmetries can
appear from non-factorisable ${\mathbb Z}_2 \times {\mathbb Z}_2$
orbifold models 
\cite{Kobayashi:2004ud,Kobayashi:2006wq}.

\subsection*{Acknowledgement}

It is a pleasure to thank Steve Abel for useful discussions.
T.~K.\/ and K.~T.\/ are supported in part by the
Grand-in-Aid for Scientific Research \#17540251 and \#172131, respectively.
T.~K.\/ is also supported in part by
the Grant-in-Aid for
the 21st Century COE ``The Center for Diversity and
Universality in Physics'' from the Ministry of Education, Culture,
Sports, Science and Technology of Japan.

\appendix

\section{Eight Classes of {\boldmath{${\mathbb Z}_2 \times {\mathbb
	Z}_2$}} Orbifolds} 

In this appendix, we list 8 classes of models. Before giving the
details of each model in subsections \ref{ap:first} -- \ref{ap:last},
we summarise some of their properties in table \ref{tab:standard}. The
Euler number, 
$\chi$, is computed according to the general formula \cite{Dixon}
\begin{equation}\label{eq:euler}
\chi = \frac{1}{\left| G \right|} \sum_{[g,h] = 0}
\chi\left(g,h\right) ,
\end{equation}
where $\left| G \right|$ is the order of the orbifold group with
elements $g$, $h$ and $\chi\left(g,h\right)$ is the number of points which are
simultaneously fixed under the action of $g$ and $h$. For the
${\mathbb Z}_2 \times {\mathbb Z}_2$ orbifold there are six
contributing pairs of non trivial orbifold group elements, each pair
leaves the same points invariant. The order of the orbifold group is
4, and hence we can simplify (\ref{eq:euler}) to
\begin{equation}
\chi = \frac{3}{2} \chi\left(\theta_1 , \theta_2\right) .
\end{equation}
For the standard embedded orbifold the net number of generations
should equal $\chi /2$. Following section \ref{sec:standard}, we can
also compute 
the number of families, $M_i$, (anti-families, $N_i$) from each
sector, i.e.\ $i=0, \ldots, 3$, where $i=0$ corresponds to the
untwisted sector. We summarise the results in table \ref{tab:standard}.

\begin{table}
\begin{center}
\begin{tabular}{| c || c| c| c| c| c| c|}
\hline 
Model &  $\chi$ & $\left( M_0 , N_0\right)$ & $\left( M_1 ,
N_1\right)$ &
$\left( M_2 , N_2\right)$ & $\left( M_3 , N_3\right)$ & $\sum_{i=0}^3
\left( M_i , N_i\right)$ \\
\hline \hline
A.1 & 96 & (3,3) & (16,0) & (16,0) & (16,0) & (51,3) \\
\hline
A.2  & 48 & (3,3) & (12,4) & (8,0) & (8,0) & (31,7) \\
\hline
A.3 & 24 & (3,3) & (10,6) & (4,0) & (4,0) & (21,9) \\
\hline
A.4 & 48 & (3,3) & (8,0) & (8,0) & (8,0) & (27,3) \\
\hline
A.5 & 24 & (3,3) & (6,2) & (6,2) & (4,0) & (19,7) \\
\hline
A.6  & 24 & (3,3) & (6,2) & (4,0) & (4,0) & (17,5) \\
\hline
A.7 & 24 & (3,3) & (4,0) & (4,0) & (4,0) & (15,3) \\
\hline
A.8 & 12 & (3,3) & (3,1) & (3,1) & (3,1) & (12,6)\\ \hline
\end{tabular}
\end{center}
\caption{The eight models for standard embedding: The numbers $M_i$ 
  ($N_i$) denote the number of (anti-)families, where the index $i$
  labels the twist sector ($i=0$ means untwisted). The net number of
  families is given by half the Euler
  number, $\chi$. \label{tab:standard}}
\end{table}   

In the following subsections we list the details for each model.

\subsection{SU(2){\boldmath{$^6$}}\label{ap:first}}

We combine six SU(2) ${\mathbb Z}_2 \times {\mathbb Z}_2$ orbifolds to 
construct 6D ${\mathbb Z}_2 \times {\mathbb Z}_2$ orbifold as follows,
\begin{center}
          \begin{tabular}{r c |c |c |c |c |c |c |c}
          & & \multicolumn{1}{c|}{$e_1$} & \multicolumn{1}{c|}{$e_2$}
            &\multicolumn{1}{c|}{$e_3$}
            & \multicolumn{1}{c|}{$e_4$} & \multicolumn{1}{c|}{$e_5$}
            & \multicolumn{1}{c|}{$e_6$} &  $N_{fp}$ \\ \hline \hline
               $\theta_1$ & : & $-$ & $-$ & $-$ & $-$ & + & +  & 16\\
               $\theta_2$ & : & + & + & $-$ & $-$ & $-$ & $-$  & 16\\
               $\theta_3$ & : & $-$ & $-$ & + & + & $-$ & $-$  & 16\\
          \end{tabular}
\end{center}
where $e_i$ denotes the simple root of the $i$-th SU(2) root
lattice, and $+$ and $-$ denote how $e_i$ transforms under
the twist $\theta_i$.
The number of fixed points (tori) for each twist is
denoted by $N_{fp}$.
At any rate, this is the conventional factorisable
6D ${\mathbb Z}_2 \times {\mathbb Z}_2$ orbifold.

%%%%%%%%%%%%%%%%%%%%%%%%%%%%%%%%%%%%%%%%%%%%%%%%%%%%%%%%%%%%%%%%%%
\subsection{SU(3){\boldmath $\times$}SU(2){\boldmath $^4$}}

We combine SU(3) ${\mathbb Z}_2 \times {\mathbb Z}_2$ orbifold and four
SU(2) ${\mathbb Z}_2 \times {\mathbb Z}_2$ orbifolds as follows,
\begin{center}

          \begin{tabular}{r c |c c |c |c |c |c|c}
          & & \multicolumn{2}{c|}{SU(3)} & \multicolumn{1}{c|}{$e_3$}
            & \multicolumn{1}{c|}{$e_5$} & \multicolumn{1}{c|}{$e_5$}
            & \multicolumn{1}{c|}{$e_6$}  & $N_{fp}$   \\ \hline \hline
               $\theta_1$ & : & $-$ & $-$ & $-$ & $-$ & $+$ & $+$ &16 \\
               $\theta_2$ & : & $+$ & $-$ & $+$ & $-$ & $-$ & $-$ & 8 \\
               $\theta_3$ & : & $-$ & $+$ & $-$ & $+$ & $-$ & $-$ & 8 \\
          \end{tabular}
\end{center}
where $e_i$ denotes the simple root of the $i$-th SU(2) root
lattice.
The twist $\theta_1$ is 2D ${\mathbb Z}_2$ rotation on the SU(3) root lattice,
and the twist $\theta_3$ the Weyl reflection corresponding to
one of SU(3) simple roots, $\alpha_1$.

%%%%%%%%%%%%%%%%%%%%%%%%%%%%%%%%%%%%%%%%%%%%%%%%%%%%%%%%%%%%%%%%%%%%%

%%%%%%%%%%%%%%%%%%%%%%%%%%%%%%%%%%%%%%%%%%%%%%%%%%%%%%%%%%%%%%%%%%
\subsection{SU(3){\boldmath $^2 \times$}SU(2){\boldmath $^2$}-I} 

We combine two SU(3) ${\mathbb Z}_2 \times {\mathbb Z}_2$ orbifolds and
two SU(2) ${\mathbb Z}_2 \times {\mathbb Z}_2$ orbifolds as follows,
\begin{center}
          \begin{tabular}{r c |c c |c c |c |c|c}
          & & \multicolumn{2}{c|}{SU(3)} & \multicolumn{2}{c|}{SU(3)}
            & \multicolumn{1}{c|}{$e_5$}  & \multicolumn{1}{c|}{$e_6$}
          & $N_{fp}$ \\ \hline \hline
               $\theta_1$ & : & $-$ & $-$ & $-$ & $-$ & $+$ & $+$ & 16 \\
               $\theta_2$ & : & $+$ & $-$ & $+$ & $-$ & $-$ & $-$ & 4 \\
               $\theta_3$ & : & $-$ & $+$ & $-$ & $+$ & $-$ & $-$ & 4 \\
          \end{tabular}
\end{center}
where the twist $\theta_1$ is 2D ${\mathbb Z}_2$ rotation on both
the first and second SU(3) lattices.
The twist $\theta_3$ is the Weyl reflection corresponding to
the first simple roots, i.e.
$\alpha_1$ and $\alpha_3$, for both the first and second
SU(3) root lattices.
%%%%%%%%%%%%%%%%%%%%%%%%%%%%%%%%%%%%%%%%%%%%%%%%%%%%%%%%%%%%%%%%%%%%%

%%%%%%%%%%%%%%%%%%%%%%%%%%%%%%%%%%%%%%%%%%%%%%%%%%%%%%%%%%%%%%%%%%
\subsection{SU(4){\boldmath  $\times$}SU(2){\boldmath $^3$}}

We combine SU(4) ${\mathbb Z}_2 \times {\mathbb Z}_2$ orbifold and three
SU(2) ${\mathbb Z}_2 \times {\mathbb Z}_2$ orbifolds as follows,
\begin{center}
          \begin{tabular}{r c |c c c |c |c |c |c}
          & & \multicolumn{3}{c|}{SU(4)} & \multicolumn{1}{c|}{$e_4$}
            & \multicolumn{1}{c|}{$e_5$} & \multicolumn{1}{c|}{$e_6$}
           &  $N_{fp}$  \\ \hline \hline
               $\theta_1$ & : & $-$ & $-$ & $+$ & $-$ & $-$ & $+$ & 8  \\
               $\theta_2$ & : & $+$ & $-$ & $-$ & $+$ & $-$ & $-$ & 8 \\
               $\theta_3$ & : & $-$ & $+$ & $-$ & $-$ & $+$ & $-$ & 8 \\
          \end{tabular}
\end{center}
where the twist $\theta_1$ is a product of the outer automorphism
replacing between $\alpha_1$ and $\alpha_3$ and the total ${\mathbb
  Z}_2$ rotation, $\alpha_i \rightarrow - \alpha_i$ ($i=1,2,3$),
for the SU(4) lattice,
while the twist $\theta_2$ is a sum of
two Weyl reflections for $\alpha_1$ and $\alpha_3$.
%%%%%%%%%%%%%%%%%%%%%%%%%%%%%%%%%%%%%%%%%%%%%%%%%%%%%%%%%%%%%%%%%%%%%

%%%%%%%%%%%%%%%%%%%%%%%%%%%%%%%%%%%%%%%%%%%%%%%%%%%%%%%%%%%%%%%%%%
\subsection{SU(3){\boldmath $^2 \times$}SU(2){\boldmath $^2$}-II}

We combine two SU(3) ${\mathbb Z}_2 \times {\mathbb Z}_2$ orbifolds and
two SU(2) ${\mathbb Z}_2 \times {\mathbb Z}_2$ orbifolds as follows,
\begin{center}
          \begin{tabular}{r c |c c |c c |c |c|c}
          & & \multicolumn{2}{c|}{SU(3)} & \multicolumn{2}{c|}{SU(3)}
            & \multicolumn{1}{c|}{$e_5$}  & \multicolumn{1}{c|}{$e_6$}
          & $ N_{fp}$\\ \hline \hline
               $\theta_1$ & : & $-$ & $-$ & $-$ & $+$ & $-$ & $+$ & 8  \\
               $\theta_2$ & : & $+$ & $-$ & $-$ & $-$ & $+$ & $-$ & 8 \\
               $\theta_3$ & : & $-$ & $+$ & $+$ & $-$ & $-$ & $-$ & 4 \\
          \end{tabular}
\end{center}
where the twist $\theta_1$ is the 2D ${\mathbb Z}_2$ rotation on
the first SU(3) lattice and the twist $\theta_3$ is
the Weyl reflection corresponding to one of simple roots of
the first SU(3) lattice, $\alpha_1$.
Moreover, the twist $\theta_1$ is the
Weyl reflection corresponding to one of simple roots of
the second SU(3) lattice, $\alpha_3$,
and the twist $\theta_2$ is the 2D ${\mathbb Z}_2$ rotation on
the second SU(3) lattice.

%%%%%%%%%%%%%%%%%%%%%%%%%%%%%%%%%%%%%%%%%%%%%%%%%%%%%%%%%%%%%%%%%%%%%

\subsection{SU(4){\boldmath $\times$}SU(3){\boldmath $\times$}SU(2)}

We combine SU(4) ${\mathbb Z}_2 \times {\mathbb Z}_2$ orbifold,
SU(3) ${\mathbb Z}_2 \times {\mathbb Z}_2$ orbifold and SU(2)
${\mathbb Z}_2 \times {\mathbb Z}_2$
orbifold as follows,
\begin{center}
          \begin{tabular}{r c |c c c |c c |c|c}
          & & \multicolumn{3}{c|}{SU(4)} & \multicolumn{2}{c|}{SU(3)}
            & \multicolumn{1}{c|}{$e_6$} & $N_{fp}$  \\ \hline \hline
               $\theta_1$ & : & $-$ & $-$ & $+$ & $-$ & $-$ & $+$ & 8  \\
               $\theta_2$ & : & $+$ & $-$ & $-$ & $+$ & $-$ & $-$ & 4 \\
               $\theta_3$ & : & $-$ & $+$ & $-$ & $-$ & $+$ & $-$ & 4 \\
          \end{tabular}
\end{center}
where the twist $\theta_1$ is a product of the outer automorphism
replacing between $\alpha_1$ and $\alpha_3$ and the total ${\mathbb
  Z}_2$ rotation, $\alpha_i \rightarrow - \alpha_i$ ($i=1,2,3$),
for the SU(4) lattice,
while the twist $\theta_2$ is a sum of
two Weyl reflections for $\alpha_1$ and $\alpha_3$.
In addition, the twist $\theta_1$ is the 2D ${\mathbb Z}_2$ rotation
on the SU(3) lattice, and the twist $\theta_3$ is
the Weyl reflection corresponding to one of SU(3)
simple roots, $\alpha_4$.

%%%%%%%%%%%%%%%%%%%%%%%%%%%%%%%%%%%%%%%%%%%%%%%%%%%%%%%%%%%%%%%%%%%%%

%%%%%%%%%%%%%%%%%%%%%%%%%%%%%%%%%%%%%%%%%%%%%%%%%%%%%%%%%%%%%%%%%%
\subsection{SU(4){\boldmath $^2$}}

We combine two SU(4) ${\mathbb Z}_2 \times {\mathbb Z}_2$ orbifolds as
follows, 
\begin{center}
          \begin{tabular}{r c |c c c |c c c|c}
          & & \multicolumn{3}{c|}{SU(4)} & \multicolumn{3}{c|}{SU(4)}
              & $N_{fp}$ \\ \hline \hline
               $\theta_1$ & : & $-$ & $-$ & $+$ & $-$ & $-$ & $+$ & 4 \\
               $\theta_2$ & : & $+$ & $-$ & $-$ & $+$ & $-$ & $-$ & 4 \\
               $\theta_3$ & : & $-$ & $+$ & $-$ & $-$ & $+$ & $-$ & 4 \\
          \end{tabular}
\end{center}
where the twist $\theta_1$ is a product of the outer automorphism
replacing between $\alpha_1$ and $\alpha_3$
($\alpha_4$ and $\alpha_6$) and the total ${\mathbb Z}_2$
rotation, $\alpha_i \rightarrow - \alpha_i$,
for the first (second) SU(4) lattice.
In addition, the twist $\theta_2$ is a sum of
two Weyl reflections for $\alpha_1$ and $\alpha_3$
($\alpha_4$ and $\alpha_6$) on the first (second) SU(4) lattice.
%%%%%%%%%%%%%%%%%%%%%%%%%%%%%%%%%%%%%%%%%%%%%%%%%%%%%%%%%%%%%%%%%%%%%

%%%%%%%%%%%%%%%%%%%%%%%%%%%%%%%%%%%%%%%%%%%%%%%%%%%%%%%%%%%%%%%%%%
\subsection{SU(3){\boldmath $^3$}} \label{ap:last}

We combine three SU(3) ${\mathbb Z}_2 \times {\mathbb Z}_2$ orbifolds
as follows, 
\begin{center}
          \begin{tabular}{r c |c c |c c |c c|c}
          & & \multicolumn{2}{c|}{SU(3)} & \multicolumn{2}{c|}{SU(3)}
            & \multicolumn{2}{c|}{SU(3)}  & $N_{fp}$ \\ \hline \hline
               $\theta_1$ & : & $-$ & $-$ & $-$ & $+$ & $-$ & $+$ & 4 \\
               $\theta_2$ & : & $+$ & $-$ & $+$ & $-$ & $-$ & $-$ & 4 \\
               $\theta_3$ & : & $-$ & $+$ & $-$ & $-$ & $+$ & $-$ & 4 \\
          \end{tabular}
\end{center}
where the twist $\theta_1$ is the 2D ${\mathbb Z}_2$ rotation on the
first SU(3) root lattice and is the Weyl reflection for one of
simple roots for both the second and third SU(3) root lattices.
The twist $\theta_3$ is the Weyl reflection for one of
simple roots for the first SU(3) root lattice and is
the 2D ${\mathbb Z}_2$ rotation on the second SU(3) root lattice.
Furthermore, the twist $\theta_2$ is the Weyl reflection
for one of simple roots on the third SU(3) root lattice.
%%%%%%%%%%%%%%%%%%%%%%%%%%%%%%%%%%%%%%%%%%%%%%%%%%%%%%%%%%%%%%%%%%%%%


\begin{thebibliography}{99}
%
\bibitem{Dixon}
L.~J.~Dixon, J.~A.~Harvey, C.~Vafa and E.~Witten, Nucl.\ Phys.\ B\
{\bf 261}, 678 (1985);
%%CITATION = NUPHA,B274,285;%%
Nucl.\ Phys.\ B\ {\bf 274}, 285 (1986).
%%CITATION = NUPHA,B261,678;%%
%
\bibitem{IMNQ}
L.~E.~Ib\'a\~nez, H.-P.~Nilles and F.~Quevedo, Phys.\ Lett.\ B\
{\bf 187}, 25 (1987);
%%CITATION = PHLTA,B187,25;%%
L.~E.~Ib\'a\~{n}ez, J.~E.~Kim, H.-P.~Nilles and F.~Quevedo, Phys.\
Lett.\ B\ {\bf 191}, 282 (1987);
%%CITATION = PHLTA,B191,282;%%
L.~E.~Ib\'a\~{n}ez, J.~Mas,
H.~P.~Nilles and F.~Quevedo, Nucl.\ Phys.\ B\ {\bf 301}, 157
(1988);
%%CITATION = NUPHA,B301,157;%%
A.~Font, L.~E.~Ib\'a\~{n}ez, F.~Quevedo and A.~Sierra,
Nucl.\ Phys.\ B\ {\bf 331}, 421 (1990);
%%CITATION = NUPHA,B331,421;%%
D.~Bailin, A.~Love and
S.~Thomas, Phys.\ Lett.\ B\ {\bf 194}, 385 (1987);
%%CITATION = PHLTA,B194,385;%%
Y.~Katsuki, Y.~Kawamura, T.~Kobayashi, N.~Ohtsubo, Y. Ono and
K.~Tanioka, Nucl.\ Phys.\ B\ {\bf 341}, 611 (1990).
%%CITATION = NUPHA,B341,611;%%
%
%\cite{Kobayashi:2004ud}
\bibitem{Kobayashi:2004ud}
T.~Kobayashi, S.~Raby and R.~J.~Zhang,
%``Constructing 5d orbifold grand unified theories from heterotic strings,''
Phys.\ Lett.\ B {\bf 593}, 262 (2004)
[arXiv:hep-ph/0403065];
%%CITATION = HEP-PH 0403065;%%
%
%
%\cite{Kobayashi:2004ya}
%\bibitem{Kobayashi:2004ya}
%T.~Kobayashi, S.~Raby and R.~J.~Zhang,
% ``Searching for realistic 4d string models with a Pati-Salam
% symmetry: Orbifold 
%grand unified theories from heterotic string compactification on a Z(6)
%orbifold,''
Nucl.\ Phys.\ B {\bf 704}, 3 (2005)
  [arXiv:hep-ph/0409098].
%%CITATION = HEP-PH 0409098;%%
%
%\cite{Forste:2004ie}
\bibitem{Forste:2004ie}
S.~F\"orste, H.~P.~Nilles, P.~K.~S.~Vaudrevange and A.~Wingerter,
%``Heterotic brane world,''
Phys.\ Rev.\ D {\bf 70}, 106008 (2004);
%[arXiv:hep-th/0406208].
%%CITATION = HEP-TH 0406208;%%
%
%
%
%\cite{Forste:2005rs}
%\bibitem{Forste:2005rs}
  S.~F\"orste, H.~P.~Nilles and A.~Wingerter,
  %``Geometry of rank reduction,''
  Phys.\ Rev.\ D {\bf 72}, 026001 (2005)
  [arXiv:hep-th/0504117];
  %%CITATION = HEP-TH 0504117;%%
%
%\cite{Forste:2005gc}
%\bibitem{Forste:2005gc}
%  S.~Forste, H.~P.~Nilles and A.~Wingerter,
  %``The Higgs mechanism in heterotic orbifolds,''
  Phys.\ Rev.\ D {\bf 73}, 066011 (2006)
  [arXiv:hep-th/0512270].
  %%CITATION = HEP-TH 0512270;%%
%


%\cite{Buchmuller:2004hv}
\bibitem{Buchmuller:2004hv}
W.~Buchm\"uller, K.~Hamaguchi, O.~Lebedev and M.~Ratz,
%``Dual models of gauge unification in various dimensions,''
 Nucl.\ Phys.\ B {\bf 712}, 139 (2005)
[arXiv:hep-ph/0412318];
%%CITATION = HEP-PH 0412318;%%
%
%\cite{Buchmuller:2005jr}
%\bibitem{Buchmuller:2005jr}
%  W.~Buchmuller, K.~Hamaguchi, O.~Lebedev and M.~Ratz,
  %``The supersymmetric standard model from the heterotic string,''
  Phys.\ Rev.\ Lett.\  {\bf 96}, 121602 (2006)
  [arXiv:hep-ph/0511035];
  %%CITATION = HEP-PH 0511035;%%
%\cite{Buchmuller:2006ik}
%\bibitem{Buchmuller:2006ik}
%  W.~Buchm\"uller, K.~Hamaguchi, O.~Lebedev and M.~Ratz,
%   ``Supersymmetric standard model from the heterotic string. II,''
  %
  arXiv:hep-th/0606187;
  %%CITATION = HEP-TH 0606187;%%
%
%\bibitem{Choi:2004wn}
  K.~S.~Choi, S.~Groot Nibbelink and M.~Trapletti,
  %``Heterotic SO(32) model building in four dimensions,''
  JHEP {\bf 0412} (2004) 063
  [arXiv:hep-th/0410232];
  %%CITATION = HEP-TH 0410232;%%
%
%\cite{Nilles:2006np}
%\bibitem{Nilles:2006np}
  H.~P.~Nilles, S.~Ramos-Sanchez, P.~K.~S.~Vaudrevange and A.~Wingerter,
  %``Exploring the SO(32) heterotic string,''
  JHEP {\bf 0604}, 050 (2006)
  [arXiv:hep-th/0603086];
  %%CITATION = HEP-TH 0603086;%%
%\bibitem{Lebedev:2006kn}
  O.~Lebedev, H.~P.~Nilles, S.~Raby, S.~Ramos-Sanchez, M.~Ratz,
  P.~K.~S.~Vaudrevange and A.~Wingerter, 
  %``A mini-landscape of exact MSSM spectra in heterotic orbifolds,''
  arXiv:hep-th/0611095,
  %%\bibitem{Lebedev:2006tr}
%  O.~Lebedev, H.~P.~Nilles, S.~Raby, S.~Ramos-Sanchez, M.~Ratz,
%  P.~K.~S.~Vaudrevange and A.~Wingerter, 
  %``Low energy supersymmetry from the heterotic landscape,''
  arXiv:hep-th/0611203;
  %%CITATION = HEP-TH 0611203;%%CITATION = HEP-TH 0611095;%%
%
%\bibitem{Kim:2006hv}
  J.~E.~Kim and B.~Kyae,
  %``String MSSM through flipped SU(5) from Z(12) orbifold,''
  arXiv:hep-th/0608085,
  %%CITATION = HEP-TH 0608085;%%
%
%\bibitem{Kim:2006hw}
%  J.~E.~Kim and B.~Kyae,
  %``Flipped SU(5) from Z(12-I) orbifold with Wilson line,''
  arXiv:hep-th/0608086.
  %%CITATION = HEP-TH 0608086;%%
%
\bibitem{Hamidi:1986vh}
  S.~Hamidi and C.~Vafa,
%   ``Interactions On Orbifolds,''
  %
  Nucl.\ Phys.\ B {\bf 279}, 465 (1987);
  %%CITATION = NUPHA,B279,465;%%
%
 %\cite{Dixon:1986qv}
\bibitem{Dixon:1986qv}
  L.~J.~Dixon, D.~Friedan, E.~J.~Martinec and S.~H.~Shenker,
  %``The Conformal Field Theory Of Orbifolds,''
  Nucl.\ Phys.\ B {\bf 282}, 13 (1987).
  %%CITATION = NUPHA,B282,13;%%
%
%\cite{Burwick:1990tu}
\bibitem{Burwick:1990tu}
T.~T.~Burwick, R.~K.~Kaiser and H.~F.~M\"uller,
%``General Yukawa Couplings Of Strings On Z(N) Orbifolds,''
Nucl.\ Phys.\ B {\bf 355}, 689 (1991);
%%CITATION = NUPHA,B355,689;%%
%
%\cite{Erler:1992gt}
%\bibitem{Erler:1992gt}
J.~Erler, D.~Jungnickel, M.~Spalinski and S.~Stieberger,
%``Higher twisted sector couplings of Z(N) orbifolds,''
Nucl.\ Phys.\ B {\bf 397}, 379 (1993).
%[arXiv:hep-th/9207049].
%%CITATION = HEP-TH 9207049;%%
%
%\cite{Kobayashi:2003vi}
\bibitem{Kobayashi:2003vi}
T.~Kobayashi and O.~Lebedev,
%``Heterotic Yukawa couplings and continuous Wilson lines,''
Phys.\ Lett.\ B {\bf 566}, 164 (2003);
%[arXiv:hep-th/0303009].
%%CITATION = HEP-TH 0303009;%%
%
%\cite{Kobayashi:2003gf}
\bibitem{Kobayashi:2003gf}
T.~Kobayashi and O.~Lebedev,
%``Heterotic string backgrounds and CP violation,''
Phys.\ Lett.\ B {\bf 565}, 193 (2003).
%[arXiv:hep-th/0304212].
%%CITATION = HEP-TH 0304212;%%
%
%\cite{Ko:2004ic}
\bibitem{Ko:2004ic}
P.~Ko, T.~Kobayashi and J.~h.~Park,
%``Quark masses and mixing angles in heterotic orbifold models,''
Phys.\ Lett.\ B {\bf 598}, 263 (2004);
%[arXiv:hep-ph/0406041].
%%CITATION = HEP-PH 0406041;%%
%\cite{Ko:2005sh}
%\bibitem{Ko:2005sh}
%  P.~Ko, T.~Kobayashi and J.~h.~Park,
  %``Lepton masses and mixing angles from heterotic orbifold models,''
  Phys.\ Rev.\ D {\bf 71}, 095010 (2005)
  [arXiv:hep-ph/0503029].
  %%CITATION = HEP-PH 0503029;%%
%
\bibitem{Antoniadis:1989zy}
%\bibitem{Kawai:1986va}
  H.~Kawai, D.~C.~Lewellen and S.~H.~H.~Tye,
  %``CONSTRUCTION OF FOUR-DIMENSIONAL FERMIONIC STRING MODELS,''
  Phys.\ Rev.\ Lett.\  {\bf 57} (1986) 1832
  [Erratum-ibid.\  {\bf 58} (1987) 429],
  %%CITATION = PRLTA,57,1832;%%  
%\bibitem{Kawai:1986ah}
%  H.~Kawai, D.~C.~Lewellen and S.~H.~H.~Tye,
  %``CONSTRUCTION OF FERMIONIC STRING MODELS IN FOUR-DIMENSIONS,''
  Nucl.\ Phys.\ B {\bf 288} (1987) 1;
  %%CITATION = NUPHA,B288,1;%%
%\bibitem{Antoniadis:1986rn}
  I.~Antoniadis, C.~P.~Bachas and C.~Kounnas,
  %``FOUR-DIMENSIONAL SUPERSTRINGS,''
  Nucl.\ Phys.\ B {\bf 289} (1987) 87;
  %%CITATION = NUPHA,B289,87;%%
%
I.~Antoniadis, J.~R.~Ellis, J.~S.~Hagelin and D.~V.~Nanopoulos,
  %``THE FLIPPED SU(5) X U(1) STRING MODEL REVAMPED,''
  Phys.\ Lett.\ B {\bf 231} (1989) 65.
  %%CITATION = PHLTA,B231,65;%%
%
%\cite{Faraggi:1991jr}
\bibitem{Faraggi:1991jr}
  A.~E.~Faraggi,
% ``A New standard - like model in the four-dimensional free fermionic
%  string formulation,''
  Phys.\ Lett.\ B {\bf 278}, 131 (1992), 
  %%CITATION = PHLTA,B278,131;%%
%
%\bibitem{Faraggi:2004rq}
  A.~E.~Faraggi, C.~Kounnas, S.~E.~M.~Nooij and J.~Rizos, 
%   ``Classification of the chiral Z(2) x Z(2) fermionic models in the
%  heterotic superstring,''
  Nucl.\ Phys.\ B {\bf 695}, 41 (2004)
  [arXiv:hep-th/0403058];
  %%CITATION = HEP-TH 0403058;%%
A.~E.~Faraggi, C.~Kounnas and J.~Rizos,
  %``Chiral family classification of fermionic Z(2) x Z(2) heterotic orbifold
  %models,''
  arXiv:hep-th/0606144;
  %%CITATION = HEP-TH 0606144;%%
A.~E.~Faraggi, E.~Manno and C.~Timirgaziu,
  %``Minimal standard heterotic string models,''
  arXiv:hep-th/0610118.
  %%CITATION = HEP-TH 0610118;%%
%
\bibitem{Faraggi:1993pr}
  A.~E.~Faraggi,
  %``Z(2) x Z(2) Orbifold compactification as the origin of realistic free
  %fermionic models,''
  Phys.\ Lett.\ B {\bf 326} (1994) 62
  [arXiv:hep-ph/9311312],
%\bibitem{Faraggi:2002qh}
%  A.~E.~Faraggi,
  %``Partition functions of NAHE-based free fermionic string models,''
  Phys.\ Lett.\ B {\bf 544} (2002) 207
  [arXiv:hep-th/0206165];
  %%CITATION = HEP-TH 0206165;%%
  %%CITATION = HEP-PH 9311312;%%
%\bibitem{Berglund:1998eq}
  P.~Berglund, J.~R.~Ellis, A.~E.~Faraggi, D.~V.~Nanopoulos and Z.~Qiu,
  %``Toward the M(F)-theory embedding of realistic free-fermion models,''
  Phys.\ Lett.\ B {\bf 433} (1998) 269
  [arXiv:hep-th/9803262].
  %%CITATION = HEP-TH 9803262;%%
%
\bibitem{Font:1988mk}
  A.~Font, L.~E.~Ib\'{a}\~{n}ez and F.~Quevedo,
  %``Z(N) x Z(M) ORBIFOLDS AND DISCRETE TORSION,''
  Phys.\ Lett.\ B {\bf 217} (1989) 272.
  %%CITATION = PHLTA,B217,272;%%
%
%\cite{Dixon:1986yc}
\bibitem{Dixon:1986yc}
  L.~J.~Dixon,
  {\it ``Symmetry Breaking in String Theories via Orbifolds,''}
PhD Thesis (1986), UMI 86-27933.
%\href{http://www.slac.stanford.edu/spires/find/hep/www?r=umi\%2F86-27933}
%{SPIRESentry}
%
\bibitem{Bailin:1994ma}
  D.~Bailin, A.~Love, W.~A.~Sabra and S.~Thomas,
  %``Modular symmetries, threshold corrections and moduli for Z(2) x Z(2)
  %orbifolds,''
  Mod.\ Phys.\ Lett.\ A {\bf 10} (1995) 337
  [arXiv:hep-th/9407049],
  %%CITATION = HEP-TH 9407049;%%
%\bibitem{Bailin:1994hu}
% D.~Bailin, A.~Love, W.~A.~Sabra and S.~Thomas,
  %``Anisotropic solutions for orbifold moduli from duality invariant gaugino
  %condensates,''
  Mod.\ Phys.\ Lett.\ A {\bf 9} (1994) 2543
  [arXiv:hep-th/9405031],
  %%CITATION = HEP-TH 9405031;%%
%\bibitem{Bailin:1993fm}
%  D.~Bailin, A.~Love, W.~A.~Sabra and S.~Thomas,
%  %``String loop threshold corrections for Z(N) Coxeter orbifolds,''
  Mod.\ Phys.\ Lett.\ A {\bf 9} (1994) 67
  [arXiv:hep-th/9310008],
  %%CITATION = HEP-TH 9310008;%%
%
%\cite{Bailin:1993ri}
%\bibitem{Bailin:1993ri}
%  D.~Bailin, A.~Love, W.~A.~Sabra and S.~Thomas,
  %``Modular Symmetries In Z(N) Orbifold Compactified String Theories With
  %Wilson Lines,''
  Mod.\ Phys.\ Lett.\ A {\bf 9} (1994) 1229
  [arXiv:hep-th/9312122],
  %%CITATION = HEP-TH 9312122;%%
%\bibitem{Bailin:1993wv}
%  D.~Bailin, A.~Love, W.~A.~Sabra and S.~Thomas,
  %``Duality Symmetries Of Threshold Corrections In Orbifold Models,''
  Phys.\ Lett.\ B {\bf 320} (1994) 21
  [arXiv:hep-th/9309133].
  %%CITATION = HEP-TH 9309133;%%
%
\bibitem{Donagi:2004ht}
  R.~Donagi and A.~E.~Faraggi,
  %``On the number of chiral generations in Z(2) x Z(2) orbifolds,''
  Nucl.\ Phys.\ B {\bf 694} (2004) 187
  [arXiv:hep-th/0403272].
  %%CITATION = HEP-TH 0403272;%%
%
\bibitem{Faraggi:2006bs}
  A.~E.~Faraggi, S.~F\"orste and C.~Timirgaziu,
  %``Z(2) x Z(2) heterotic orbifold models of non factorisable six dimensional
  %toroidal manifolds,''
  JHEP {\bf 0608} (2006) 057
  [arXiv:hep-th/0605117].
  %%CITATION = HEP-TH 0605117;%%
%
\bibitem{Font:1988mm}
  A.~Font, L.~E.~Ib\'{a}\~{n}ez, H.~P.~Nilles and F.~Quevedo,
  %``Yukawa Couplings In Degenerate Orbifolds: Towards A Realistic
  %SU(3) X SU(2) 
  %X U(1) Superstring,''
  Phys.\ Lett.\  {\bf 210B} (1988) 101
  [Erratum-ibid.\ B {\bf 213} (1988) 564].
  %%CITATION = PHLTA,210B,101;%% 
%\cite{Font:1988tp}
%\bibitem{Font:1988tp}
%  A.~Font, L.~E.~Ibanez, H.~P.~Nilles and F.~Quevedo,
  %``Degenerate Orbifolds,''
  Nucl.\ Phys.\ B {\bf 307} (1988) 109
  [Erratum-ibid.\ B {\bf 310} (1988) 764].
  %%CITATION = NUPHA,B307,109;%%
%%%%%%%%%%%%%%%%%%%%%%%%%%%%%%%%%%%%%%%%%%%%%%%%%%%%%%%%%%%%%%%
\bibitem{Narain:1986qm}
  K.~S.~Narain, M.~H.~Sarmadi and C.~Vafa,
  %``ASYMMETRIC ORBIFOLDS,''
  Nucl.\ Phys.\ B {\bf 288} (1987) 551.
  %%CITATION = NUPHA,B288,551;%%
%
 \bibitem{Erler:1992ki}
  J.~Erler and A.~Klemm,
  %``Comment On The Generation Number In Orbifold Compactifications,''
  Commun.\ Math.\ Phys.\  {\bf 153} (1993) 579
  [arXiv:hep-th/9207111].
  %%CITATION = HEP-TH 9207111;%%
%
\bibitem{Vafa:1994rv}
  C.~Vafa and E.~Witten,
  %``On orbifolds with discrete torsion,''
  J.\ Geom.\ Phys.\  {\bf 15} (1995) 189
  [arXiv:hep-th/9409188].
  %%CITATION = HEP-TH 9409188;%%
%
\bibitem{Kobayashi:1990mi}
  T.~Kobayashi and N.~Ohtsubo,
  %``Analysis on the Wilson lines of Z(N) orbifold models,''
  Phys.\ Lett.\ B {\bf 257}, 56 (1991).
  %%CITATION = PHLTA,B257,56;%%
%
%\cite{Kobayashi:1991rp}
\bibitem{Kobayashi:1991rp}
  T.~Kobayashi and N.~Ohtsubo,
  %``Geometrical aspects of Z(N) orbifold phenomenology,''
  Int.\ J.\ Mod.\ Phys.\ A {\bf 9}, 87 (1994).
  %%CITATION = IMPAE,A9,87;%%
%
%\cite{Kobayashi:2006wq}
\bibitem{Kobayashi:2006wq}
  T.~Kobayashi, H.~P.~Nilles, F.~Pl\"oger, S.~Raby and M.~Ratz,
  %``Stringy origin of non-Abelian discrete flavor symmetries,''
  arXiv:hep-ph/0611020.
  %%CITATION = HEP-PH 0611020;%%
\end{thebibliography}
\end{document}